\documentclass[lettersize,preprint]{IEEEtran}
\usepackage{amsmath,amsfonts}
\usepackage{algorithmic}
\usepackage{algorithm}
\usepackage{footnote}
\makesavenoteenv{algorithm}
\makesavenoteenv{algorithmic}
\usepackage{array}
\usepackage[caption=false,font=normalsize,labelfont=sf,textfont=sf]{subfig}
\usepackage{textcomp}
\usepackage{stfloats}
\usepackage{url}
\usepackage[hidelinks]{hyperref}
\usepackage{verbatim}
\usepackage{graphicx}
\usepackage{cite}
\usepackage{amsthm}
\usepackage{bm} 
\usepackage{xcolor}
\usepackage{orcidlink}
\usepackage{flushend}  

\usepackage{tikz}
\usetikzlibrary{positioning}
\usepackage{tkz-graph}
\usepackage{pgfplots}
\usepackage{listofitems}
\usetikzlibrary{math, calc, positioning, shapes, backgrounds, fit, arrows, decorations.pathreplacing, graphs, spy}
\pgfplotsset{compat=newest}

\begin{document}

\title{DMD Prediction of MIMO Channel Using Tucker Decomposition}


\author{Irina Kopnina\,\orcidlink{0009-0007-5649-7799}, Dmitry Artemasov\,\orcidlink{0000-0003-2054-2846}, Sergey Matveev\,\orcidlink{0000-0001-8000-8595}
\thanks{Irina Kopnina and Sergey Matveev are with the Lomonosov MSU, Faculty of Computational Mathematics and Cybernetics, Moscow, Russia and Marchuk Institute of Numerical Mathematics RAS, Moscow, Russia;  Dmitry Artemasov is with the Center for Next Generation Wireless and IoT (NGW), Skolkovo Institute of Science and Technology, Moscow, Russia (emails: kopninaia@my.msu.ru, d.artemasov@skoltech.ru, matseralex@cs.msu.ru).}
\thanks{This work was supported by the Russian Science Foundation (project No. 25-11-00392).}
}


\maketitle

\begin{abstract}

Accurate channel state information (CSI) prediction is crucial for next-generation multiple-input multiple-output (MIMO) communication systems. Classical prediction methods often become inefficient for high-dimensional and rapidly time-varying channels. To improve prediction efficiency, it is essential to exploit the inherent low-rank tensor structure of the MIMO channel. Motivated by this observation, we propose a dynamic mode decomposition (DMD)–based prediction framework operating on the low-dimensional core tensors obtained via a Tucker decomposition. The proposed method predicts reduced-order channel cores, significantly lowering computational complexity. Simulation results demonstrate that the proposed approach preserves the dominant channel dynamics and achieves high prediction accuracy.

\end{abstract}


\vspace{-0.5cm}
\section{Introduction}

\IEEEPARstart{A}{ccurate} channel state information (CSI) is essential for multiple-input multiple-output (MIMO) systems, enabling spatial multiplexing and beamforming gains. In wideband orthogonal frequency-division multiplexing (OFDM) systems, CSI is required across all subcarriers; however, user mobility and Doppler effects cause channel aging, leading to significant performance degradation \cite{Truong2013ChannelAging}.

To mitigate channel aging, a range of channel prediction techniques has been investigated in the literature, including parametric model-based approaches such as autoregressive (AR) modeling~\cite{Baddour2005AR}, spatio-temporal autoregressive (ST-AR) models~\cite{Wu2021STAR}, and Prony-based prediction in the angular–delay domain~\cite{Yin2020Prony}. However, multipath propagation and Doppler effect can limit the accuracy of these classical methods, leading to degraded performance in mobile scenarios. More recently, data-driven and dynamical-systems-based predictors have attracted increasing attention. Machine learning methods~\cite{Xiao2024ML,Kim2021ML} and deep learning approaches~\cite{Li2024DL,Jiang2019} can capture underlying channel structure, but they often incur high computational complexity and require large training datasets, which limits their practical adoption.

Dynamic mode decomposition (DMD) is a data-driven technique for modeling the temporal evolution of high-dimensional systems using a linear dynamical approximation. It was introduced in~\cite{DMDbook} and originally applied to fluid dynamics prediction. Later, it was adopted for approximating the dominant channel dynamics from a small number of CSI observations, as well as for CSI feedback and channel prediction~\cite{HaddadICC2023DMD}, with extensions incorporating adaptive quantization~\cite{Zhu2024RVQDMD}, compressed sensing~\cite{Haddad2024CSDMD}, and convolutional autoencoders~\cite{Haddad2025DMDCAE}.

In parallel, the high-dimensional nature of MIMO-OFDM channels has motivated tensor-based channel modeling, in which the channel is characterized jointly across spatial, frequency, and temporal dimensions. Tensor-based channel processing has been widely studied. Canonical polyadic decomposition (CPD) and Tucker decomposition have been applied to channel estimation and denoising tasks~\cite{Zhou2017Tensor,Amin2017HOSVD}. More recent works have employed Tucker-structured models to capture spatial--frequency--temporal correlations and non-stationarity~\cite{Hou2024ADD,Li2026Tucker}, while tensor formulations combined with DMD enable multidimensional harmonic retrieval and channel prediction~\cite{Zhang2024TensorDMD}. Furthermore, continuous-time prediction using tensor neural ordinary differential equations (TN-ODEs) has been proposed to avoid interpolation errors by modeling intra-frame channel evolution~\cite{Cui2023TNODE}. However, most existing approaches neglect the inherent low-rank structure of the channel and instead attempt to predict the entire channel tensor directly.

To address this challenge, we propose a DMD--Tucker--based channel prediction framework that models the temporal evolution of low-dimensional channel cores, followed by reconstruction of the full channel using slowly varying orthogonal factor matrices. The key idea is to compute the Tucker decomposition only for the initial channel observation, while the core tensors of subsequent observations are obtained via orthogonal projections onto the fixed factor matrices. This significantly reduces computational complexity while yielding the same reduced DMD operator as predicting the full channel directly.

The rest of the paper is organized as follows. Section~\ref{sec:system_model} introduces the system model and formulates the channel prediction problem. Section~\ref{sec:tucker_channel} describes the Tucker representation of the channel. Section~\ref{sec:dmd_channel} describes the application of DMD to channel prediction using the full channel tensor. Section~\ref{sec:dmd_tucker} presents the proposed Tucker-based core prediction approach and proves that the reduced linear operator learned by DMD is equivalent for both full-tensor and core-based representations. Numerical results evaluating the performance of the proposed DMD--Tucker method are provided in Section~\ref{sec:results}.

\section{System Model} \label{sec:system_model}

In this paper, we consider a time division duplex (TDD) MIMO-OFDM wireless communication system.

By $N_{\mathrm{rx}}$, $N_{\mathrm{tx}}$, and $N_{\mathrm{sc}}$ we denote the number of user equipment (UE) receive antennas, base station (BS) transmit antennas, and subcarriers, respectively. At each discrete time slot $t$, the frequency-domain MIMO channel is represented by a third-order tensor
\begin{equation}
\mathcal{H}_t \in \mathbb{C}^{N_{\mathrm{rx}}\times N_{\mathrm{tx}} \times N_{\mathrm{sc}}},
\end{equation}
where the $(i,j,k)$-th element of $\mathcal{H}_t$ corresponds to the complex channel gain between the $i$-th receive antenna and the $j$-th transmit antenna at the $k$-th subcarrier.

We assume that channel realizations are available over $T$ consecutive time slots, forming a temporal sequence $\{ \mathcal{H}_1, \mathcal{H}_2, \ldots, \mathcal{H}_T \}$. The objective of this work is to exploit the inherent low-rank multilinear structure of $\mathcal{H}_t$ and the underlying temporal dynamics to predict future channel tensors $\mathcal{H}_{t+\tau}$ for $\tau > 0$.

Mathematically the problem of channel prediction at future time slots can be formulated as 
\begin{equation}
\min_{\mathcal{\widehat{H}}_{t+\tau}}
\left\|
\mathcal{H}_{t+\tau} - \mathcal{\widehat{H}}_{t+\tau}
\right\|_F^2, \
\end{equation}
where $H_{t+\tau}$ represents the true CSI and $\mathcal{\widehat{H}}_{t+\tau}$ represents the predicted CSI at $t+\tau$-th time slot.

\section{Tucker Representation of the Channel}\label{sec:tucker_channel}
Practical MIMO-OFDM wireless channels exhibit strong correlations across spatial and frequency domains due to limited scattering, antenna geometry, and finite delay spread. Such correlations imply that the channel tensor admits a low multilinear rank representation, which has been widely exploited in recent works on channel estimation and modeling~\cite{Artemasov2024ItuVAR,Sidiropoulos2017,Cheng2018,Yuan2020, petrov2024}.

To capture this structure, the channel tensor at time slot $t$, $\mathcal{H}_t \in \mathbb{C}^{N_{\mathrm{rx}} \times N_{\mathrm{tx}} \times N_{\mathrm{sc}}}$, is approximated using the Tucker decomposition as
\begin{equation} \label{eq:tucker}
\mathcal{H}_t \approx \mathcal{G}_t 
\times_1 \mathbf{U}_{\mathrm{rx}_{t}} 
\times_2 \mathbf{U}_{\mathrm{tx}_{t}} 
\times_3 \mathbf{U}_{\mathrm{sc}_{t}},
\end{equation}
where $\times_n$ denotes the mode-$n$ tensor--matrix product. The factor matrices
$\mathbf{U}_{\mathrm{rx}_{t}} \in \mathbb{C}^{N_{\mathrm{rx}} \times R_{\mathrm{rx}}}$,
$\mathbf{U}_{\mathrm{tx}_{t}} \in \mathbb{C}^{N_{\mathrm{tx}} \times R_{\mathrm{tx}}}$, and
$\mathbf{U}_{\mathrm{sc}_{t}} \in \mathbb{C}^{N_{\mathrm{sc}} \times R_{\mathrm{sc}}}$
span the dominant subspaces associated with the receive antennas, transmit antennas, and subcarriers, respectively. In such notation, $R_{\mathrm{rx}}$, $R_{\mathrm{tx}}$, $R_{\mathrm{sc}}$ define the ranks of the orthogonal matrices. The core tensor
$\mathcal{G}_t \in \mathbb{C}^{R_{\mathrm{rx}} \times R_{\mathrm{tx}} \times R_{\mathrm{sc}}}$
encodes the coupling among these low-dimensional subspaces.

Moreover, the factor matrices are assumed to vary slowly over time, reflecting quasi-static propagation characteristics. 
Thus, channel dynamics are mainly captured by the core tensor $\mathcal{G}_t$, making Tucker decomposition suitable for channel prediction in a reduced-dimensional space~\cite{Park2019}.

\section{DMD-Based Channel Prediction} \label{sec:dmd_channel}
\subsection{Prediction of Full Tensor }
In this section, we describe the application of classical DMD for channel prediction by learning the temporal dynamics of the vectorized CSI tensors.
We denote the vectorized CSI tensor $\mathcal{H}_t$ as

\begin{equation}
\mathbf{h}_t = \mathrm{vec}(\mathcal{H}_t) \in \mathbb{C}^{N},
\end{equation}
where $N = N_{\mathrm{rx}} N_{\mathrm{tx}} N_{\mathrm{sc}}$. By stacking $T$ consecutive channel snapshots, we construct two one-step shifted matrices
\begin{equation}
\mathbf{X} = [\mathbf{h}_1, \mathbf{h}_2, \ldots, \mathbf{h}_{T-1}], \quad
\mathbf{Y} = [\mathbf{h}_2, \mathbf{h}_3, \ldots, \mathbf{h}_T].
\end{equation}

DMD assumes the existence of a linear operator $\mathbf{A}$ such that $ \mathbf{Y} \approx \mathbf{A}\mathbf{X}$,
where $\mathbf{A}$ captures the underlying temporal dynamics of the channel evolution. The optimal operator in the least-squares sense is given by
$\mathbf{A} = \mathbf{Y}\mathbf{X}^{\dagger}$
with $(\cdot)^{\dagger}$ denoting the Moore--Penrose pseudoinverse.  

To reduce computational complexity, a truncated singular value decomposition (SVD) of $\mathbf{X}$ is computed as $
\mathbf{X} \approx \mathbf{U}_r \boldsymbol{\Sigma}_r \mathbf{V}_r^{H},
$ 

where $r$ is the selected rank. The reduced-order DMD operator is then obtained as
\begin{equation}
\tilde{\mathbf{A}} = \mathbf{U}_r^{H} \mathbf{Y} \mathbf{V}_r \boldsymbol{\Sigma}_r^{-1}.
\end{equation}
The eigen-decomposition of $\tilde{\mathbf{A}}$ yields $\tilde{\mathbf{A}}\mathbf{W} = \mathbf{W}\boldsymbol{\Lambda},$

where $\boldsymbol{\Lambda}$ contains the DMD eigenvalues and $\mathbf{W}$ contains the corresponding eigenvectors. The DMD modes are given by $\boldsymbol{\Phi} = \mathbf{Y}\mathbf{V}_r \boldsymbol{\Sigma}_r^{-1} \mathbf{W}. $

Using the identified DMD modes and eigenvalues, the channel vector at a future time slot $t+\tau$ is predicted as
\begin{equation}\label{ed:pred_next}
\hat{\mathbf{h}}_{t+\tau} = \boldsymbol{\Phi}\boldsymbol{\Lambda}^{\tau}\mathbf{b},
\end{equation}
where the coefficient vector $\mathbf{b} = \mathbf{\Phi^{\dagger}} \mathbf{h}_1$. Finally, the predicted channel vector is reshaped to obtain the predicted channel tensor
\begin{equation}
\hat{\mathcal{H}}_{t+\tau} = \mathrm{unvec}(\hat{\mathbf{h}}_{t+\tau}).
\end{equation}

While this vectorized DMD approach captures temporal channel dynamics, it ignores the inherent multilinear structure of the wireless channel and suffers from high computational complexity when the channel dimension is large. These limitations motivate the Tucker-based DMD prediction framework proposed in the next section.

\section{Tucker  Channel Predicition}\label{sec:dmd_tucker}
\subsection{Tucker  Channel Predicition Algorithm}

The proposed approach is inspired by~\cite{Kutz2016}, where DMD is applied to data after an orthogonal transformation, exploiting the invariance of the reduced DMD operator under isometric mappings.
In this work, we extend this principle to a multilinear setting by employing Tucker decomposition for CSI modeling.

The key assumption underlying the Tucker-based formulation is that the dominant spatial and spectral subspaces associated with the receiver antennas, transmitter antennas, and subcarriers vary slowly over time: $\mathbf{U}_{\mathrm{rx}_{1}} \approx \mathbf{U}_{\mathrm{rx}_{t}}$, $\mathbf{U}_{\mathrm{tx}_{1}} \approx \mathbf{U}_{\mathrm{tx}_{t}}$, $\mathbf{U}_{\mathrm{sc}_{1}} \approx \mathbf{U}_{\mathrm{sc}_{t}}$.

After computing the Tucker decomposition for the first CSI tensor, the core tensors corresponding to subsequent channel snapshots are obtained via orthogonal projections onto these fixed subspaces as
\begin{equation}
\mathcal{G}_{t}
=
\mathcal{H}_{t}
\times_1 \mathbf{U}^{H}_{\mathrm{rx}}
\times_2 \mathbf{U}^{H}_{\mathrm{tx}}
\times_3 \mathbf{U}^{H}_{\mathrm{sc}},
\qquad \forall t\in[1,\dots,T] .
\end{equation}

In Algorithm \ref{alg:tucker_dmd}, DMD-Tucker prediction is described. This formulation enables efficient temporal modeling and prediction using DMD in the reduced-dimensional core space while preserving the essential channel dynamics.

\begin{algorithm}[t]
\caption{Tucker--DMD Prediction of MIMO Channel}
\label{alg:tucker_dmd}
\begin{algorithmic}[1]

\REQUIRE Tensor $\mathcal{H}_1 \in \mathbb{C}^{N_{\mathrm{\mathrm{rx}}} \times N_{\mathrm{\mathrm{tx}}} \times N_{\mathrm{\mathrm{sc}}}}$, time samples $\{\mathcal{H}_{t} \}_{t=1}^{T}$
\ENSURE Predicted channel $\widehat{\mathcal{H}}_{T+\tau}$, $\forall \tau > 0$

\STATE \textbf{Tucker decomposition of first channel:}
\vspace{-6pt}
\[
\mathcal{H}_1 \approx \mathcal{G}_1 \times_1 \mathbf{U}_{\mathrm{rx}_1} \times_2 \mathbf{U}_{\mathrm{tx}_1} \times_3 \mathbf{U}_{\mathrm{sc}_1}\vspace{-6pt}
\]

\STATE \textbf{Form core trajectories over time:}
\vspace{-6pt}
\[
\mathcal{G}_{t} = \mathcal{H}_{t} \times_1 \mathbf{U}^{H}_{\mathrm{rx}_t} \times_2 \mathbf{U}^{H}_{\mathrm{tx}_t} \times_3 \mathbf{U}^{H}_{\mathrm{sc}_t}{,\:\forall t\in[1,\dots,T]} \vspace{-6pt}
\]

\STATE \textbf{Stack cores for DMD:}
\vspace{-6pt}
\[
\mathbf{X_G} = [\,\mathbf{g}_1, \mathbf{g}_2,\dots,\mathbf{g}_{T-1}\,], ~~
\mathbf{Y_G} = [\,\mathbf{g}_2, \mathbf{g}_3,\dots,\mathbf{g}_{T}\,], \vspace{-6pt}
\]
where $\mathbf{g}_t = \mathrm{vec}(\mathcal{G}_{t}) \in \mathbb{C}^{r_1 r_2 r_3}$.

\STATE \textbf{DMD step for cores} using $\mathbf{X_G}$ and $\mathbf{Y_G}$ : 
$\widehat{\mathbf{g}}_{T+\tau} = \boldsymbol{\Phi}\boldsymbol{\Lambda}^{\tau}\mathbf{b}$,
where $\mathbf{b} = \mathbf{\Phi^{\dagger}} \mathbf{g}_1$.

\STATE \textbf{Reshape predicted core:}
\vspace{-6pt}
\[
\widehat{\mathcal{G}}_{T+\tau} = \mathrm{unvec}\!\left(\widehat{\mathbf{g}}_{T+\tau}\right) \vspace{-17pt}
\]
\STATE \textbf{Reconstruct full predicted tensor:}
\vspace{-4pt}
\[
\widehat{\mathcal{H}}_{T+\tau} =
\widehat{\mathcal{G}}_{T+\tau}
\times_1 \mathbf{U}_{\mathrm{rx}_1} \times_2 \mathbf{U}_{\mathrm{tx}_1} \times_3 \mathbf{U}_{\mathrm{sc}_1}
\]

\RETURN $\widehat{\mathcal{H}}_{T+\tau}$
\end{algorithmic}
\end{algorithm}

\subsection{Similar eigenvalues for the Full DMD and Tucker DMD}

Vectorizing the Tucker decomposition \eqref{eq:tucker} yields the standard Kronecker form
\begin{equation}
\mathbf{h}_t := \mathrm{vec}(\mathcal{H}_t)
\;\approx\;
\left(\mathbf{U}_{\mathrm{sc}_{1}} \otimes \mathbf{U}_{\mathrm{tx}_{1}} \otimes \mathbf{U}_{\mathrm{rx}_{1}}\right)\, \mathrm{vec}(\mathcal{G}_t)
\;=:\; \mathbf{C}\,\mathbf{g}_t,
\label{eq:vec_relation}
\end{equation}
where $\mathbf{g}_t := \mathrm{vec}(\mathcal{G}_t)$ and
\[
\mathbf{C} := \mathbf{U}_{\mathrm{sc}_1} \otimes \mathbf{U}_{\mathrm{tx}_1} \otimes \mathbf{U}_{\mathrm{rx}_1} \in \mathbb{C}^{(N_{\mathrm{sc}} N_{\mathrm{tx}} N_{\mathrm{rx}})\times (R_{\mathrm{sc}} R_{\mathrm{tx}} R_{\mathrm{rx}})}.
\]
Because each $\mathbf{U}_{\mathrm{sc}_1} , \mathbf{U}_{\mathrm{tx}_1} , \mathbf{U}_{\mathrm{rx}_1} $ has orthonormal columns, $\mathbf{C}$ is an isometry:
\begin{equation}
\mathbf{C}^H \mathbf{C} = \mathbf{I}_{R_{\mathrm{sc}} R_{\mathrm{tx}} R_{\mathrm{rx}}}.
\label{eq:T_isometry}
\end{equation}

\paragraph{Snapshot matrices}
Given a sequence $\{\mathbf{h}_t\}_{t=1}^{m}$ and $\{\mathbf{g}_t\}_{t=1}^{m}$ related by \eqref{eq:vec_relation}, define the standard DMD snapshot matrices
\[
\mathbf{X}_H := [\mathbf{h}_1,\dots,\mathbf{h}_{m-1}],\qquad
\mathbf{Y}_H := [\mathbf{h}_2,\dots,\mathbf{h}_m],
\]
and analogously in the core space
\[
\mathbf{X}_G := [\mathbf{g}_1,\dots,\mathbf{g}_{m-1}],\qquad
\mathbf{Y}_G := [\mathbf{g}_2,\dots,\mathbf{g}_m].
\]
Stacking \eqref{eq:vec_relation} across time gives the induced relation
\begin{equation}
\mathbf{X}_H \approx \mathbf{C}\,\mathbf{X}_G,\qquad
\mathbf{Y}_H \approx \mathbf{C}\,\mathbf{Y}_G.
\label{eq:snapshot_relation}
\end{equation}

\paragraph{Relation between SVD of cores and full channel}
Let us assume that truncated SVD of $\mathbf{X}_G$ is given by
\begin{equation}
\mathbf{X_G} = \mathbf{U_G}\,\mathbf{\Sigma_G}\,\mathbf{V}^H_{\mathbf{G}},
\label{eq:svd_XG}
\end{equation}

Using \eqref{eq:snapshot_relation}, we obtain SVD of $\mathbf{X}_H$ 
\begin{equation}
\mathbf{X}_H \approx \mathbf{C}\,\mathbf{X}_G = (\mathbf{C}\mathbf{U_G})\,\mathbf{\Sigma_G}\,\mathbf{V_G}^H.
\label{eq:svd_lifted}
\end{equation}
By \eqref{eq:T_isometry} and $\mathbf{U_G}^H \mathbf{U_G} = \mathbf{I}$, the lifted left factor $\mathbf{U}_H := \mathbf{C}\mathbf{U_G}$ also has orthonormal columns:
\[
\mathbf{U}_H^H \mathbf{U}_H = \mathbf{U_G}^H (\mathbf{C}^H \mathbf{C}) \mathbf{U_G} = \mathbf{I}.
\]
Hence \eqref{eq:svd_lifted} is an SVD up to the same truncation level and approximation error. Singular values and right singular vectors are the same in $\mathbf{X_G}$ and $\mathbf{X_H}$ : $\mathbf{\Sigma_H} = \mathbf{\Sigma_G}$, $\mathbf{V_H} = \mathbf{V_G}$.

\paragraph{Reduced DMD operators coincide}
In (projected) DMD, the reduced operator associated with $(\mathbf{X}, \mathbf{Y})$ is
\[
\widetilde{\mathbf{A}} := \mathbf{U}^H \mathbf{Y} \mathbf{V}\,\mathbf{\Sigma}^{-1},
\]
where $\mathbf{X} = \mathbf{U}\mathbf{\Sigma}\mathbf{V}^H$ is the chosen (possibly truncated) SVD.
Applying this to the core snapshots gives
\begin{equation}
\widetilde{\mathbf{A}}_{\mathbf{G}} = \mathbf{U_G}^H \mathbf{Y_G} \mathbf{V_G}\,\mathbf{\Sigma_G}^{-1}.
\label{eq:AG_tilde}
\end{equation}
For the full snapshots, using $\mathbf{X_H} \approx (\mathbf{C}\mathbf{U_G})\mathbf{\Sigma_G}\mathbf{V_G}^H$ and $\mathbf{Y_H} \approx \mathbf{C}\mathbf{Y_G}$ yields
\begin{align}
\widetilde{\mathbf{A}}_H
&:= \mathbf{U}^H_{\mathbf{H}} \mathbf{Y_H} \mathbf{V_H}\,\mathbf{\Sigma_H}^{-1}
= (\mathbf{C}\mathbf{U_G})^H (\mathbf{C}\mathbf{Y_G}) \mathbf{V_G}\,\mathbf{\Sigma}^{-1}_{\mathbf{G}} \nonumber\\
&= \mathbf{U_G}^H (\mathbf{C}^H \mathbf{C}) \mathbf{Y_G} \mathbf{V_G}\,\mathbf{\Sigma}^{-1}_{\mathbf{G}}
= \mathbf{U}^H_{\mathbf{G}} \mathbf{Y_G} \mathbf{V_G}\,\mathbf{\Sigma}^{-1}_{\mathbf{G}}
= \widetilde{\mathbf{A}}_G.
\label{eq:AH_equals_AG}
\end{align}
Therefore, the reduced DMD operators computed in the full vectorized space and in the Tucker core space are \emph{identical} $\widetilde{\mathbf{A}}_{\mathbf{H}} = \widetilde{\mathbf{A}}_{\mathbf{G}}$ (under exact relations \eqref{eq:snapshot_relation} and orthonormality \eqref{eq:T_isometry}).
Note that the equivalence between the full and Tucker-based formulations holds only when the Tucker decomposition \eqref{eq:tucker} approximates the original data with sufficiently small Frobenius-norm error. If it is large, the channel vector cannot be accurately represented as in (\ref{eq:vec_relation}), and the resulting operators are no longer equal. 

\section{Results}\label{sec:results}
 
\subsection{Simulation Scenario}

\begin{table}[!t]
    \caption{Main simulation parameters}
    \centering
    \begin{tabular}{|c|c|}
    \hline
    Parameter type                    & \multicolumn{1}{c|}{Parameter value} \\
    \hline\hline
    Channel model type                 & \multicolumn{1}{c|}{\begin{tabular}{@{}c@{}}3GPP 38.901 UMa NLoS \cite{3GPP901} \end{tabular}} \\ 
    \hline
    \begin{tabular}{@{}c@{}}Antenna configuration \\ {[UE/BS]} \end{tabular} & \multicolumn{1}{c|}{4/64} \\
    \hline
    Bandwidth, MHz                     & \multicolumn{1}{c|}{100} \\ 
    \hline
    Subcarrier spacing, kHz            & \multicolumn{1}{c|}{30} \\
    \hline
    Number of pilot subcarriers~\cite{3GPP211}          & \multicolumn{1}{c|}{1632} \\ 
    \hline
    SRS period $T_p$, ms                     & \multicolumn{1}{c|}{{5\dots 20}} \\
    \hline
    Carrier frequency, GHz             & \multicolumn{1}{c|}{3.5} \\ 
    \hline
    Distance to UEs, m                 & \multicolumn{1}{c|}{50}     \\
    \hline    
    UEs speed, km/h                    & \multicolumn{1}{c|}{5 } \\
    \hline
    \end{tabular}
    \label{tab:scenario}
\end{table}
We consider the 3GPP-compliant time-varying channel simulation, with the main parameters defined in the Table~\ref{tab:scenario}.
The proposed algorithm is compared with well-established baselines and state-of-the-art tensor-based prediction methods:
\begin{itemize}
    \item \textbf{Zero-order-hold (ZOH):} 
    Assumes a fixed channel over the prediction horizon: $\mathcal{H}_{T+\tau} = \mathcal{H}_{T}, \forall \tau > 0$.
    \item \textbf{AR~\cite{Baddour2005AR}:} 
    Each channel coefficient is predicted as a linear combination of its past samples.\footnote{We note that AR multi-step forecasts must be computed recursively, using each predicted value as input to the next step. In our approach, multi-step forecasts follow directly by recomputing the core representation using \eqref{ed:pred_next}, with no need to rerun the algorithm.}
    \item \textbf{Tucker--AR (T-AR)~\cite{TuckerAR}:}     
    AR is employed to predict the core, while the orthogonal factor matrices are estimated once and kept fixed over time, as defined in~\cite[eq. 5]{TuckerAR}
    \item \textbf{Full-channel DMD~\cite{HaddadICC2023DMD}:} 
    Applies DMD directly to the full-dimensional channel to extrapolate future states, as described in Section~\ref{sec:dmd_channel}.
    \item \textbf{Proposed Tucker-DMD (T-DMD):} 
    As described in Section~\ref{sec:dmd_tucker}, the channel is projected onto fixed orthogonal factors obtained from the initial observation, DMD is applied to the resulting core coefficients for temporal prediction, and future channels are reconstructed using the predicted cores and the fixed factors.
\end{itemize}

\subsection{Experiments}

Prediction accuracy is measured by the NMSE,
\begin{equation}
\mathrm{NMSE} = \frac{||\mathcal{H}_{T+\tau} - \mathcal{\widehat{H}}_{T+\tau}||_F}{||\mathcal{H}_{T+\tau}||_F}.
\end{equation}

To obtain reliable estimates, we report NMSE values averaged over 1000 Monte Carlo simulations, which reduces statistical variability. For all algorithms except ZOH, predictions are computed using the last $10$ historical snapshots.

The original tensor $\mathcal{H}_1 \in \mathbb{C}^{4 \times 64 \times 1632}$ is compressed using Tucker decomposition for T-DMD and T-AR algorithms. The multilinear ranks are selected adaptively by truncating mode-wise  singular values below a prescribed relative threshold\footnote{The threshold is inherently data-dependent, as the scale and decay of singular values can vary across datasets. Therefore, the threshold used here was selected empirically for the considered data.}: $\varepsilon_1 = 10^{-5}$ and $\varepsilon_2 = 10^{-3}$.  While Tucker decomposition can be performed with predefined multilinear ranks, such an approach requires prior knowledge of the intrinsic tensor structure. Since this information is unavailable in advance, ranks are selected adaptively using an accuracy-based truncation threshold.

\begin{figure}[t]
    \centering
    \includegraphics{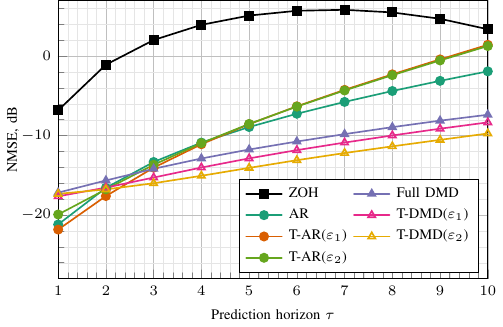}
    \vspace{-10pt}
    \caption{Channel prediction performance for different prediction horizons with a fixed channel measurement period ${T_p=5\:\mathrm{ms}}$ and under SNR = 30 dB}
    \label{fig:per_tti}
\end{figure}

\figurename~\ref{fig:per_tti} provides the results of the NMSE performance in relation to the prediction horizon $\tau$. In this experiment, we consider SNR $=30$ dB reception\footnote{While this setting may not be fully representative of practical operating conditions, it provides a controlled low-noise scenario that allows us to assess the method’s performance when noise contamination is relatively small.}, and the fixed channel measurement period $T_p=5$\,ms. The results indicate that the AR model provides accurate short-term predictions but lacks robustness for long-term forecasting. In contrast, DMD-based methods can capture channel dynamics over a larger prediction horizon without the need to recalculate the estimation. T-DMD maintains an NMSE level of approximately $-10$ dB up to the tenth prediction time slot with $\varepsilon_2$, whereas AR-based methods achieve comparable accuracy only up to the fourth prediction time slot. For Tucker-based methods using a threshold of $\varepsilon_1 = 10^{-5}$ results in a core tensor
$\mathcal{G}_1 \in \mathbb{C}^{4 \times 64 \times 256}$,
corresponding to a compression ratio of approximately $6.4\times$
while preserving high reconstruction quality. Increasing the truncation threshold to $10^{-3}$ leads to a more aggressive rank reduction,
yielding a core tensor
$\mathcal{G}_1 \in \mathbb{C}^{4 \times 37 \times 25}$,
and a compression ratio of approximately $112.9\times$. We note that increasing $\varepsilon_{2}$ further would not result in additional denoising; rather, it would remove useful information, i.e., it would discard signal content instead of suppressing noise.

\begin{figure}[t]
    \centering
    \includegraphics{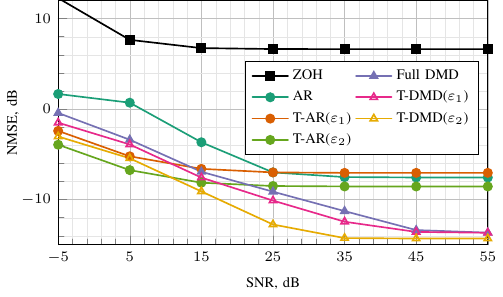}
    \vspace{-10pt}
    \centering\caption{Channel prediction performance for different SNR values at a fixed prediction horizon $\tau=5$ and a fixed channel measurement period $T_p=5\:\mathrm{ms}$}
    \label{fig:per_snr}
\end{figure}

\figurename~\ref{fig:per_snr} depicts the results of channel estimation performance in relation to the SNR at $\tau=5$. In the low-SNR regime ($-5$ to $5$ dB), Tucker--AR achieves the best NMSE due to the inherent robustness of AR modeling, where temporal averaging reduces sensitivity to noise. DMD-based methods remain competitive 
\begin{figure}[H]
    \centering
    \vspace{-10pt}
    \includegraphics{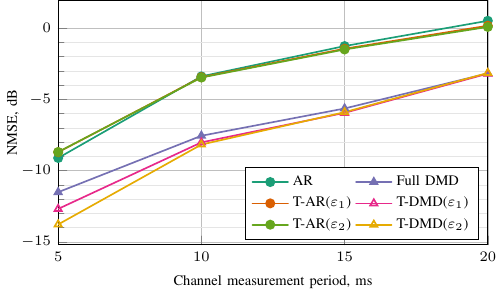}
    \vspace{-10pt}
    \centering
    \caption{Channel prediction performance for different channel measurement periods under noiseless conditions at a fixed prediction horizon, $\tau=5$}
    \label{fig:per_pilot_period}
\end{figure}
\noindent
but are slightly degraded since noise directly affects the estimation of the global dynamical operator, and the resulting modeling errors propagate during prediction. Note that due to low SNR, algorithms with different $\varepsilon_1$ and $\varepsilon_2$ have almost the same compression rate.  As the SNR increases, Full-DMD and T-DMD significantly outperform AR-based approaches because the underlying dynamics can be estimated more accurately. AR models remain limited by their local temporal structure, whereas DMD captures global spatio-temporal dynamics. At high SNR, Full-DMD and T-DMD exhibit nearly identical performance, since both rely on the same dynamical operator \eqref{eq:AH_equals_AG} and the Tucker decomposition preserves the dominant signal subspace with negligible approximation error, yielding equivalent accuracy with reduced computational complexity.

\figurename~\ref{fig:per_pilot_period} analyzes the performance of the considered methods in relation to the pilot transmission period under fixed prediction horizon $\tau=5$. This analysis illustrates the robustness of the considered methods under increasingly fast channel dynamics. The results show that DMD-based approaches consistently outperform AR-based methods for all measurement periods, indicating their superior ability to capture and extrapolate the underlying channel dynamics even when the channel changes rapidly between measurements.
\section{Conclusion}

In this paper, a novel Tucker-DMD approach for channel prediction is proposed. It is shown that the reduced dynamical operator obtained in the Tucker representation is equivalent to the operator of the full DMD formulation, up to the approximation error. The proposed method is evaluated under realistic channel simulations, demonstrating the effectiveness of Tucker-DMD for accurate and stable long-term channel prediction while preserving the predictive capability of full DMD in a reduced-dimensional space.

Future work will focus on a detailed complexity analysis of the Tucker-DMD framework and on extending the approach using advanced DMD methods, including higher-order DMD (HODMD) and optimized DMD (OptDMD).

 \bibliographystyle{IEEEtran}
\bibliography{ref}
\newpage

\vfill

\end{document}